\documentclass[
prc,%
10pt,%
final,%
notitlepage,%
oneside,%
twocolumn,%
nobibnotes,%
nofootinbib,
superscriptaddress,%
floatfix,%
showkeys,%
showpacs]%
{revtex4-1}

\usepackage{color} 
\usepackage{amsfonts} 
\usepackage{amsbsy} 
\usepackage{mathrsfs}
\usepackage{graphicx}
\usepackage{comment}

\def\lsim{\mathrel{\rlap{ \lower4pt\hbox{\hskip-3pt$\sim$}}
    \raise1pt\hbox{$<$}}} 
\def\gsim{\mathrel{\rlap{ \lower4pt\hbox{\hskip-3pt$\sim$}}
    \raise1pt\hbox{$>$}}} 
\def\scr#1{\mbox{\scriptsize #1}}

\begin{document}

\title{Update of the Three-fluid Hydrodynamics-based Event Simulator: \\
light-nuclei production in heavy-ion collisions} 


%
\author{M. Kozhevnikova}\thanks{e-mail: kozhevnikova@jinr.ru}
\affiliation{Veksler and Baldin Laboratory of High Energy Physics,
  JINR Dubna, 141980 Dubna, Russia}
\author{Yu. B. Ivanov}\thanks{e-mail: yivanov@theor.jinr.ru}
\affiliation{Bogoliubov Laboratory of Theoretical Physics, JINR Dubna,
  141980 Dubna, Russia} 
  \affiliation{National Research Nuclear
  University "MEPhI",
  115409 Moscow, Russia} 
  \affiliation{National Research Centre
  "Kurchatov Institute", 123182 Moscow, Russia}
\author{Iu. Karpenko}\thanks{e-mail: yu.karpenko@gmail.com}
\affiliation{Faculty of Nuclear Sciences and Physical Engineering, Czech Technical University in Prague,\\ 
11519 Prague 1, Czech Republic}
\author{D. Blaschke}\thanks{e-mail: blaschke@ift.uni.wroc.pl}
\affiliation{Institute of Theoretical Physics, University of Wroclaw,
  50-204 Wroclaw, Poland} 
\affiliation{Bogoliubov Laboratory of
  Theoretical Physics, JINR Dubna, 141980 Dubna, Russia}
\affiliation{National Research Nuclear University "MEPhI",
  115409 Moscow, Russia}
\author{O. Rogachevsky}\thanks{e-mail: rogachevsky@jinr.ru}
\affiliation{Veksler and Baldin Laboratory of High Energy Physics,
  JINR Dubna, 141980 Dubna, Russia}

\begin{abstract}
We present an update of the event generator based on the three-fluid dynamics (3FD), complemented by Ultra-relativistic Quantum Molecular Dynamics (UrQMD) for the late stage of the nuclear collision~-- the 
Three-fluid Hydrodynamics-based Event Simulator Extended 
by UrQMD final State interactions (THESEUS). 
Two modifications are introduced. 
The THESEUS table of hadronic resonances is made consistent with that of the underlying 3FD model. 
The main modification is that the generator is extended to 
simulate the light-nuclei production in relativistic heavy-ion collisions, on the equal basis with hadrons. These modifications are illustrated 
by applications to the description of available experimental data. 
The first run of the updated generator revealed a  good reproduction of 
the NA49 data on the light nuclei. The reproduction is achieved 
without any extra parameters, while the coalescence approach in 3FD requires special tuning of the coalescence
coefficients for each light nucleus separately. 
\end{abstract}
%
%
\maketitle

\section{Introduction}

In Ref.~\cite{Batyuk:2016qmb} 
the THESEUS event generator
was presented and its possible applications to the description of 
heavy-ion collisions were demonstrated. 
The THESEUS is based on the 3FD model~\cite{Ivanov:2005yw,Ivanov:2013wha}
extended by UrQMD~\cite{Bass:1993em,Bass:1998ca} for the afterburner stage.
The 3FD  was designed to simulate heavy-ion collisions at moderately relativistic energies 
of the Beam Energy Scan program (BES) at the Relativistic Heavy-Ion Collider~(RHIC)
at the Brookhaven National Laboratory~(BNL)~\cite{Stephans:2006tg},
CERN Super-Proton-Synchrotron~(SPS)~\cite{SPS-scan}, 
the Facility for Antiproton and Ion Research~(FAIR) in Darmstadt~\cite{FAIR} and the
Nuclotron-based Ion Collider fAcility~(NICA) in Dubna~\cite{NICA}.
Precisely in this energy range the onset of deconfinement is expected \cite{Ivanov:2013wha}.

The 3FD approximation is a minimal way to simulate the
early, nonequilibrium stage of the produced strongly-interacting matter.
It takes into account counterstreaming of the leading baryon-rich matter at the early stage of
nuclear collisions \cite{Ivanov:2005yw}. 
This nonequilibrium stage is
modeled by the means of two counterstreaming baryon-rich fluids, which are initially
associated with the constituent nucleons of the projectile (p) and the target (t) nuclei.  
Later, these fluids may consist of any type of constituents
rather than only nucleons.  
Newly produced particles, which dominantly populate the midrapidity region,
are associated with a fireball~(f) fluid.  Each of these fluids is
governed by conventional hydrodynamic equations coupled by friction
terms in the right-hand sides of the Euler equations.  These friction
terms describe the energy--momentum exchange between the fluids.

Different equations of state (EoS) can be applied within the 3FD
model.  The recent series of 3FD simulations
were based on three different types of the EoS: a purely
hadronic EoS \cite{gasEOS} (hadr. EoS) and two versions of the EoS
with deconfinement \cite{Toneev06}, i.e. 
an EoS with a first-order phase transition (1PT EoS) and one with a
smooth crossover transition (crossover EoS).  Analysis of available
experimental data indicated a strong preference of the deconfinement
scenarios with the deconfinement already at moderately relativistic
energies $\sqrt{s_{NN}}>$ 5 GeV~\cite{Ivanov:2013yqa}.

When the system becomes dilute, the 3FD evolution is stopped and
the system is frozen out.  The local freeze-out criterion used in the
3FD model is $\varepsilon < \varepsilon_{\scr{frz}}$, where~$\varepsilon$ is the local energy density of all three fluids in their common rest frame and 
$\varepsilon_{\scr{frz}}=$ 0.4 GeV/fm$^3$ is the freeze-out energy density serving as a freeze-out criterion. The value of $\varepsilon_{\scr{frz}}$ is a so-called trigger quantity, at which the analysis of other freeze-out conditions starts. The actual freeze-out occurs at lower energy density that depends on these additional freeze-out checks.  
More details can be found in Refs.~\cite{71,74}. 
The output of the model is recorded in terms of
Lagrangian test particles (i.e. fluid droplets) for each fluid
$\alpha$ (= p, t or~f), which are characterized by a local flow velocity and thermodynamic quantities.

The original 3FD model still needs a certain refinement. 
An afterburner stage that can play
an important role for some observables is absent in the model.   
From the practical point of
view, the model is not well suited for data simulations in terms of
experimental events, because the model output consists of fluid
characteristics rather than of a set of observable particles
since the 3FD observables are computed directly via the integrals over the freeze-out hypersurface.

The event generator THESEUS developed in Ref.~\cite{Batyuk:2016qmb}
solves both the above-mentioned problems.  It presents
the 3FD output in terms of a set of observed particles and the
afterburner is incorporated by means of the UrQMD model
\cite{Bass:1993em,Bass:1998ca}. 
Since the time THESEUS was first presented, certain updates have been made which we describe in this paper. The updated THESEUS we further refer to as THESEUS-v2.

\section{Rapidity distributions} 
  \label{Rapidity distributions}

In THESEUS-v2, we have updated the list of hadronic resonances used for hadron sampling to be identical  to the list of hadronic resonances in the underlying 3FD model. The updated hadron list is shorter as compared to the initial version of THESEUS.
The hadron list now includes only hadrons with well-known decay modes \cite{pdg2005}. 
This list is quite sufficient for the moderately high collision energies, for which the 
3FD model is designed. At such energies, the relative contribution from highly excited resonances to the yields of stable hadrons is quite small. The reason is a lower temperature at the surface of fluid-to-particle transition, as compared to the LHC energies. A detailed discussion of this issue is presented in Appendix \ref{Resonances}.

This update requires 
THESEUS-v2 to be validated against the underlying 3FD  model. In this section we do this check 
at the example of the Au+Au collisions at the collision energy of $\sqrt{s_{NN}}=$ 9.2 GeV.  
This example is chosen also because there are  experimental data \cite{Abelev:2009bw}
for this reaction and these data were not analyzed within neither the THESEUS, nor the 3FD before. 
Apart from methodological purposes, the analysis of this Au+Au reaction will show us 
the effect of the UrQMD afterburner stage on various species produced in the reaction. 

\begin{figure*}[!tbh]
  \includegraphics[width=.95\textwidth]{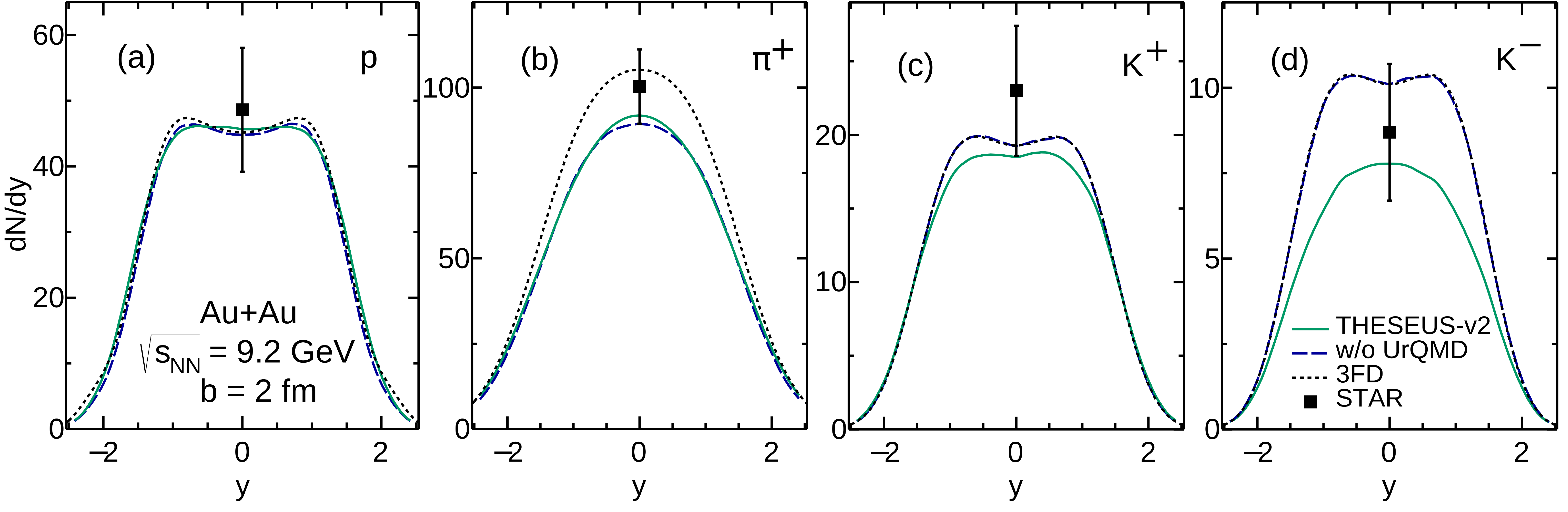}
  \caption{
Rapidity distribution of protons~(a), of positive pions~(b) 
(excluding the 
weak-decay feed-down),
of positive kaons~(c) and
of negative kaons~(d)
in central ($b=$~2~fm) Au+Au collisions
at a collision energy of $\sqrt{s_{NN}}=$ 9.2~GeV 
calculated with the crossover EoS \cite{Toneev06}: 
the 3FD result (short-dashed line), 
 the result of THESEUS-v2  
without UrQMD (long-dashed line) and
the result of THESEUS-v2 with UrQMD (solid line). 
Experimental data are from the STAR collaboration  \cite{Abelev:2009bw}.
}	
    \label{fig:Au43b2}
\end{figure*}

Fig.~\ref{fig:Au43b2}~(a) demonstrates the proton rapidity distribution.
The weak-decay protons are included, consistently with the
STAR data point \cite{Abelev:2009bw}, which is overlaid on the plot.
The 3FD table of hadronic resonances (\mbox{THESEUS-v2} without UrQMD) leads to a good 
reproduction of the 3FD result. The effect of the UrQMD afterburner for protons turns out 
to be small, which one can see from `THESEUS-v2` curve in Fig.~\ref{fig:Au43b2}~(a).

Similar results for pions are demonstrated in Fig.~\ref{fig:Au43b2}~(b). 
The weak-decay contributions to the pion spectra are excluded, to be again consistent with the STAR data~\cite{Abelev:2009bw}. 
As seen from Fig.~\ref{fig:Au43b2}~(b), the THESEUS-v2 result differs from the 
3FD one in spite of the identical lists of resonances. This is a consequence of different 
decay branching fractions of hadronic resonances in the \mbox{THESEUS-v2} and 3FD codes.
The data on branching fractions collected by the Particle Data Group~(PDG) for their Review of Particle Physics, are corrected in each edition thereof. Moreover, the branching ratios are experimentally known with an accuracy of~20--40\%, as a rule, even for well-established decay modes.
This concerns also the probabilities of decays with one and two final pions, which are of prime importance 
for the rapidity distributions. 
Therefore, implementations of the branching data differ among different models. In practice, the decay channels and
their branching fractions for THESEUS are taken from the EPOS3 code~\cite{Werner:2013tya}. Therefore, somewhat different 
decay channels the THESEUS-v2 and 3FD codes result in the difference displayed 
in~Fig.~\ref{fig:Au43b2}~(b). We have checked that the thermal pion distributions are identical in 
the THESEUS-v2 and 3FD calculations. 
Moreover, the UrQMD afterburner only slightly changes the rapidity distribution of pions, similarly to that of protons.

The results for positive and negative kaons are presented in Fig.~\ref{fig:Au43b2}~(c) and~(d), respectively. 
THESEUS-v2 without UrQMD perfectly reproduces the 3FD results. 
The UrQMD afterburner stronger affects kaons than protons and pions.
In particular, it strongly reduces the $K^-$ midrapidity density because of strong absorption of $K^-$ in reactions of the type $K^- + n \to \Lambda + \pi^{-} $.
In particular, this strong reduction of $K^-$ solves the problem of overestimation of the 
$K^-$ yield in the 3FD \cite{Ivanov:2013yqa}. 

\section{Light Nuclei}
\label{nuclei}

An important modification consists in inclusion of light nuclei in the list of particles: 
deuterons ($d$), tritons ($t$), helium isotopes $^3$He and $^4$He, 
and low-lying resonances of the $^4$He system, the decays of which contribute to the yields 
of stable species  \cite{Shuryak:2019ikv}, see Tab.\ \ref{tab:clusters}. 
\begin{table}[ht]
\begin{center}
\begin{tabular}{|c|c|c|}
\hline
Nucleus($E$[MeV]) & $J$ &  decay modes, in \% \\
\hline
\hline
$d$           & $1$ & Stable \\
$t$           & $1/2$ & Stable \\
$^3$He        & $1/2$ & Stable \\
$^4$He        & $0$ & Stable \\
$^4$He(20.21) & $0$ & $p$ = 100\\
$^4$He(21.01) & $0$ & $n$ = 24,  $p$ = 76\\
$^4$He(21.84) & $2$ & $n$ = 37,  $p$ = 63  \\
$^4$He(23.33) & $2$ & $n$ = 47,  $p$ = 53  \\
$^4$He(23.64) & $1$ & $n$ = 45,  $p$ = 55 \\
$^4$He(24.25) & $1$ & $n$ = 47,  $p$ = 50,  $d$ = 3 \\
$^4$He(25.28) & $0$ & $n$ = 48,  $p$ = 52\\
$^4$He(25.95) & $1$ & $n$ = 48,  $p$ = 52 \\
$^4$He(27.42) & $2$ & $n$ = 3,   $p$ = 3,   $d$ = 94 \\
$^4$He(28.31) & $1$ & $n$ = 47,  $p$ = 48,  $d$ = 5 \\
$^4$He(28.37) & $1$ & $n$ = 2,   $p$ = 2,   $d$ = 96 \\
$^4$He(28.39) & $2$ & $n$ = 0.2, $p$ = 0.2, $d$ = 99.6 \\
$^4$He(28.64) & $0$ & $d$ = 100 \\
$^4$He(28.67) & $2$ & $d$ = 100 \\
$^4$He(29.89) & $2$ & $n$ = 0.4, $p$ = 0.4, $d$ = 99.2 \\
\hline
\end{tabular}
\caption{Stable light nuclei and low-lying resonances of the~$^4$He system (from BNL properties of nuclides.\footnote{
\url{https://www.nndc.bnl.gov/nudat2/getdataset.jsp?nucleus=4HE&unc=nds}}). 
$J$ denotes the total angular momentum. 
The last column represents branching ratios of the decay channels, in per cents. The 
$p,n,d$ correspond to the emission of proton, neutron, or deuteron, respectively.
}
\label{tab:clusters}
\end{center}
\end{table}
The momentum distributions of these 
nuclei are sampled similarly to those of other hadrons, i.e.\ according to their 
phase-space distribution functions at given flow velocity, 
chemical potentials and temperature. 
Contrary to other hadrons, the light nuclei do not participate in the UrQMD afterburner, 
 because the UrQMD is unable to propagate them. 
However, the above sampling cannot be done straightforward, proceeding from the 3FD input.

The original 3FD model calculates spectra of the so-called primordial nucleons, i.e. both observable 
nucleons and those bound in the light nuclei. 
Therefore, the nucleons bound in the light nuclei should be 
subtracted from the primordial ones in order to obtain the observable 
nucleons which can be compared with data. 
Such subtraction is performed in the 3FD, where the spectra of the light nuclei are calculated within the coalescence approach \cite{Ivanov:2005yw,Ivanov:2017nae} 
rather than the statistical one.
\mbox{THESEUS} takes temperature and chemical potential fields for hadron sampling from the hydrodynamic evolution in the 3FD (where the clusters are not included in the EoS), and produces both hadrons and clusters with the statistical approach. This leads to an overestimate of the total baryon charge of the final state hadrons+clusters, therefore a compensating correction has to be made.
Such correction is made by means of the recalculation of 
the baryon chemical potential, proceeding from the local
baryon number conservation the ensemble of hadrons extended by
the light-nuclei species listed in Tab.\ \ref{tab:clusters}.
Details of this recalculation are described in Appendix \ref{Recalculation}.

\begin{figure*}[!tbh]
  \includegraphics[width=.95\textwidth]{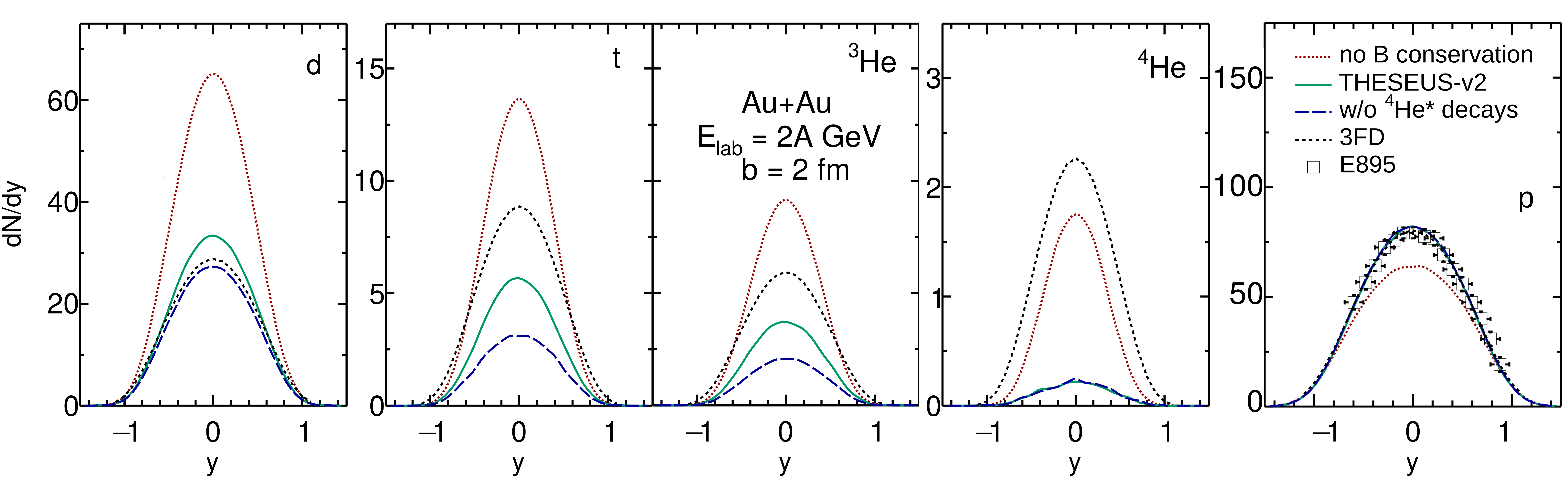}
  \caption{
Rapidity distributions of deuterons ($d$), tritons ($t$), $^3$He, $^4$He and protons in central ($b=$ 2 fm) Au+Au collisions
at  collision energy of 
$\sqrt{s_{NN}}=$ 2.7 GeV ($E_{\rm lab}=2A$~GeV)
within the crossover scenario \cite{Toneev06} calculated by means of: 
the 3FD with coalescence (short-dashed line), the THESEUS-v2 without recalculation of the baryon chemical potential and hence without baryon charge conservation 
(dotted line), THESEUS-v2 without excited states of $^4$He (long-dashed line), 
and  full THESEUS-v2 (solid line).
The proton data are from Ref. \cite{Klay:2003zf}.
}
\label{fig:Au2mix}
\end{figure*}

\begin{figure*}[!tbh]
  \includegraphics[width=.95\textwidth]{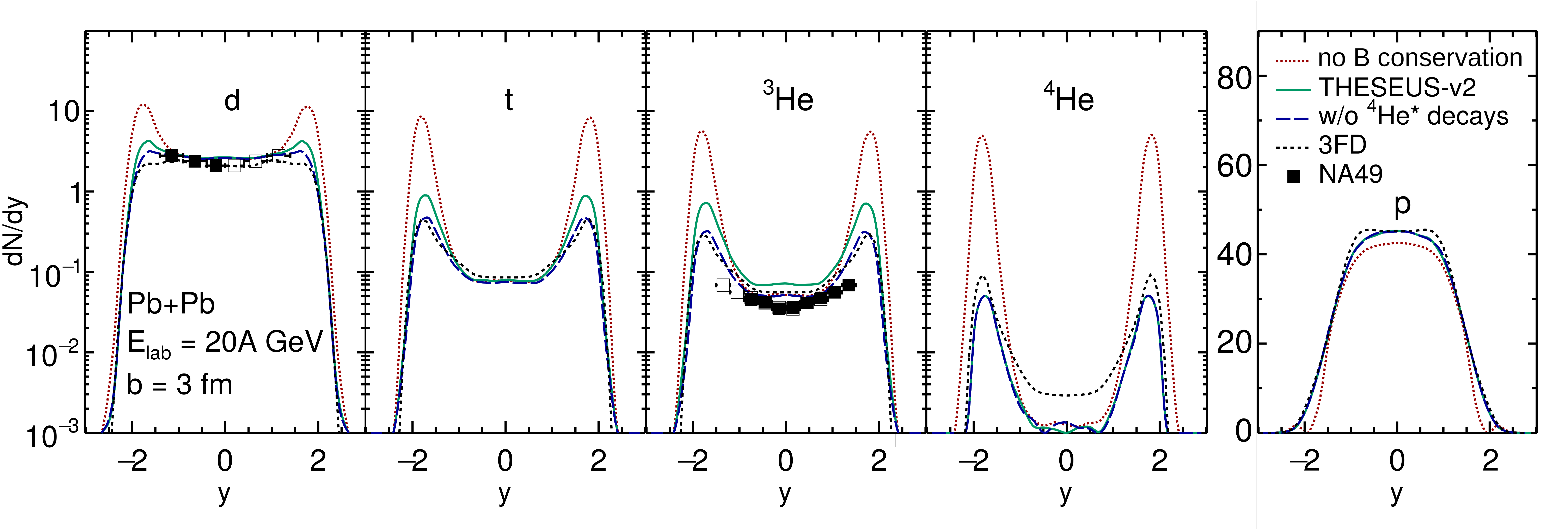}
  \caption{
The same as in Fig. \ref{fig:Au2mix} but for central ($b=$ 3 fm) 
Pb+Pb collisions at a collision energy of $E_{\rm lab}=$ 20$A$ GeV. 
Experimental data are from the NA49 collaboration  \cite{Anticic:2016ckv}.
}	
    \label{fig:Pb20mix_i6}
\end{figure*}

In hydrodynamic simulations at top RHIC or LHC energies, nucleon clusters typically do not enter in  the equation of state of the hadronic phase. Even if they did, the clusters would have a negligible impact on the thermodynamic quantities. The reason is that at low baryon chemical potential, yields of clusters are strongly suppressed due to their mass. At lower collision energies, large baryon chemical potential partially conpensates for the difference between the masses of baryons and baryonic clusters; in the statistical hadronization approach, production of each new cluster state is  suppressed by a factor of $\exp(-(m_N-\mu_B)/T)$ as compared to $\exp(-m_N/T)$ at zero baryon chemical potential.

Therefore, below we also demonstrate results without the above recalculation of 
the baryon chemical potential in order to indicate the collision energies and kinematic 
regions, where such recalculation is vital.  
The generated  tritons and $^3$He mainly differ by their isotopic content. 
Appendix \ref{Isotopic} describes the way of treatment of the isotopic content 
within the 3FD and THESEUS-v2 simulations. 
For comparison, we also present results of the 3FD coalescence \cite{Ivanov:2005yw,Ivanov:2017nae}. 
An overview of the coalescence is presented in Appendix \ref{Coalescence}.

Results of such simulations within the THESEUS-v2 are demonstrated in Fig. \ref{fig:Au2mix} for central~($b=$ 2 fm) Au+Au collisions 
at a collision energy of~$\sqrt{s_{NN}}=$~2.7~GeV 
($E_{\rm lab}=2$A~GeV in notation of E895 \cite{Klay:2003zf})
in the crossover EoS scenario, which corresponds to the curve labeled `THESEUS-v2`.
In order to illustrate the effect of  excited states of~$^4$He, 
we also show results without their contributions. 
As one can see, the recalculation of the baryon chemical potential leads to a considerable reduction of the light-nuclei yields,  cf. THESEUS-v2 and "no~$B$ conservation" curves in Fig. \ref{fig:Au2mix}. 
The inclusion of the excited states of $^4$He also noticeably affects the 
light-nuclei yields.

The results of the 3FD calculation within the coalescence approach are also presented in Fig. \ref{fig:Au2mix}. 
As the light-nuclei data for this reaction are absent, the coalescence coefficients in the 3FD were chosen 
solely on the condition that the proton result agrees with the E895 data \cite{Klay:2003zf} 
after subtraction of the light-nuclei contribution from the yield of the primordial protons. 
In practice, the coalescence coefficients are taken from Ref.\ \cite{Russkikh:1993ct} and are scaled 
 to reproduce the observed proton rapidity distribution, with their ratios fixed, see Appendix \ref{Coalescence}. 
Therefore, a detailed comparison of THESEUS-v2 results with those of the 3FD is meaningful only for protons. 
Although the experimental data on the light nuclei are unavailable, 
we can conclude that the light-nuclei yield predicted by THESEUS-v2 
is reasonable because the resulting proton rapidity distribution agrees well with the E895 data \cite{Klay:2003zf}. 
The agreement is achieved without any fitting (or extra) parameters, contrary to the coalescence approach in the 3FD.

Figure \ref{fig:Pb20mix_i6} shows similar results at a higher collision energy of 
$\sqrt{s_{NN}}=$~6.4~GeV ($E_{\rm lab}=20$A~GeV in notation of NA49 \cite{Anticic:2016ckv}) 
for central~($b=$~3~fm) Pb+Pb collisions.  
The shown protons do not include contributions from the weak decays. 
These results are confronted to the experimental data on the light-nuclei production
by the NA49 collaboration  \cite{Anticic:2016ckv}. 
The impact parameter of the simulation 
corresponds to the centrality selection of the data.  
The reproduction of the data turns out to be quite good. 
Here the multiplicities of produced light nuclei are considerably lower than those at the low energy 
considered above.
The readjustment of the baryon chemical potential strongly affects this result as well,
but only at the forward and backward rapidities. 
At the same time, the effect of the excited states of $^4$He becomes weak. 
The results of the 3FD coalescence calculation from Ref.\ \cite{Ivanov:2017nae} are shown as well.
It is worthwhile to mention that the THESEUS-v2 simulation reproduces the data \cite{Anticic:2016ckv} well
without any fit parameters, whereas the 3FD coalescence requires a special tuning of the coalescence
coefficients for each light nucleus separately.

\section{Summary}
\label{Summary}

An update of the event generator THESEUS~-- THESEUS-v2 is presented.
We benchmark the update in Au-Au collisions at $\sqrt{s_{NN}}=9.2$ and~2.7~GeV and Pb-Pb collisions at~$\sqrt{s_{NN}}=6.4$~GeV.

The first modification concerns the THESEUS table of hadronic resonances, which is made consistent with that of 
the underlying 3FD model. 
It is done in order to exactly conserve the energy, strangeness and baryon number.
The effect of the resonance table is stronger at higher collision energies because a larger number of hadronic resonances beyond the 
3FD list are noticeably excited. 
At $\sqrt{s_{NN}}<$ 5 GeV the effect of the resonance table
is negligible.

The production of light nuclei, i.e.\ deuterons~($d$), tritons~($t$), helium isotopes~$^3$He and~$^4$He, and low-lying resonances of the~$^4$He, is implemented in the generator.
The low-lying $^4$He resonances decay into nucleons and lighter nuclei, thereby contributing to their total yields.
The yields and momenta of the light nuclei are sampled in the thermodynamic approach, similarly to those of hadrons,
i.e.\ according to the local equilibrium distribution functions with a given flow velocity, temperature and chemical potentials.
Note that within the 3FD the spectra 
of the light nuclei are calculated within the coalescence
approach rather than the thermodynamic one. 
Therefore, a recalculation of the baryonic chemical potentials, 
provided by the 3FD output, was performed. 
It was demonstrated that the recalculation is very important for the 
light-nuclei production in the thermodynamic approach,  
especially at low collision energies and at the forward and backward rapidities 
in the high-energy collisions. 
The effect of the decays of low-lying resonances of the~$^4$He is large at low collision energies and gradually dies out with 
increasing energy. 

The first application of THESEUS-v2 revealed a good reproduction of the NA49 data~\cite{Anticic:2016ckv} on light nuclei. This reproduction was achieved without any extra (fit) parameters, while the 3FD coalescence~\cite{Ivanov:2017nae} required a tuning of the coalescence coefficients for each light nucleus separately. 
In principle, the coalescence coefficients can be calculated proceeding from  
phase-space and quantum mechanical aspects of nuclei formation \cite{Scheibl:1998tk}.

At present, there are several 3D dynamical models which include the coalescence mechanism of the light-nuclei production 
\cite{Ivanov:2005yw,Ivanov:2017nae,Liu:2019nii,Zhu:2015voa,Dong:2018cye,Sombun:2018yqh,Zhao:2020irc},
see also a recent reviews \cite{Oliinychenko:2020ply,Donigus:2020fon}. 
The recently developed transport models 
SMASH (Simulating Many Accelerated Strongly-interacting Hadrons)     
\cite{Weil:2016zrk,Oliinychenko:2018ugs} and PHQMD (Parton-Hadron-Quantum-Molecular-Dynamics) 
\cite{Aichelin:2019tnk}, treat light nuclei microscopically 
(so far, only deuterons in the SMASH \cite{Oliinychenko:2018ugs}) 
on an equal footing with other hadrons. 
THESEUS-v2 incorporates the thermodynamic approach in the 3FD model, which is an alternative to the above approaches.
We expect that all these models will serve as complementary approaches to study the light-nuclei production at collision energies of the BES-RHIC, SPS, NICA and FAIR.
 

\begin{acknowledgments}

This work was partially supported by the Russian Foundation for Basic Research (RFBR) 
under Grant Nos. 18-02-40084, 18-02-40085 (Y.B.I.) and 18-02-40137 (D.B.).  
D.B., Y.B.I., M.K. and O.R. acknowledge support through the 
Russian Science Foundation under project No. 17-12-01427.
D.B. and Y.B.I. received support from the Ministry of Education and Science of the Russian Federation within the Academic Excellence Project of the 
NRNU MEPhI under contract No. 02.A03.21.0005.  
I.K. acknowledges support by the project Centre of Advanced Applied Sciences with number CZ.02.1.01/0.0/0.0/16-019/0000778, which is co-financed by the European Union.

\end{acknowledgments}

\appendix
\section{Table of Hadronic Resonances} 
  \label{Resonances}

The table of hadronic resonances incorporated 
into the original THESEUS-v1 model~\cite{Batyuk:2016qmb} is presented in Tab.~\ref{tab:hadrons}.
This list of hadronic resonances is much longer than that the one incorporated in the 3FD. 
The 3FD list includes hadrons with well-known decay modes \cite{pdg2005}.
The hadronic states in the 3FD are marked with bold font in~Tab.~\ref{tab:hadrons}.
Only hadrons are displayed in Tab. \ref{tab:hadrons}, the 
corresponding anti-hadrons (where applicable) are implied. 
Different isotopic states of the listed hadrons are also implied 
but not explicitly displayed. 
\begin{table}[!ht]
{
  \begin{tabular}{l|c|c|c}
    \hline \hline
    light  &   &   &      \\
    unflavored  & flavored  & N and $\Delta$  & flavored     \\
     mesons &  mesons &  baryons &  baryons    \\
    \hline
   \pmb{$\pi$}           &         \pmb{$K$}             &  \pmb{$N$}            &  \pmb{$\Lambda$}\\
   \pmb{$\eta$}          &         $K_0^*$(800)          &  \pmb{$N$(1440)}      &  $\Lambda$(1405)\\
   \pmb{$f_0$(600)}      &         \pmb{$K^*$(892)}      &  \pmb{$N$(1520)}      &  $\Lambda$(1520)\\
   \pmb{$\rho$(770)}     &         $K_1$(1270)           &  \pmb{$N$(1535)}      &  $\Lambda$(1600)\\
   \pmb{$\omega$(782)}   &         $K_1$(1400)           &  \pmb{$N$(1650)}      &  $\Lambda$(1670)\\
   \pmb{$\eta'$(958)}    &         $K^*$(1410)           &  \pmb{$N$(1675)}      &  $\Lambda$(1690)\\
   \pmb{$f_0$(980)}      &         $K_0^*$(1430)         &  \pmb{$N$(1680)}      &  $\Lambda$(1800)\\
   \pmb{$a_0$(980)}      &         $K_2^*$(1430)         &  \pmb{$N$(1700)}      &  $\Lambda$(1810)\\
   \pmb{$\phi$(1020)}    &         $K$(1460)             &  \pmb{$N$(1710)}      &  $\Lambda$(1820)\\
   $h_1$(1170)     &         $K_2$(1580)                 &  \pmb{$N$(1720)}      &  $\Lambda$(1830)\\
   $b_1$(1235)     &         $K_1$(1650)                 &  $N$(2190)            &  $\Lambda$(1890)\\
   $a_1$(1260)     &         $K^*$(1680)                 &  \pmb{$\Delta$(1232)} &  $\Lambda$(2100)\\
   $f_2$(1270)     &         $K_2$(1770)                 &  \pmb{$\Delta$(1600)} &  $\Lambda$(2110)\\
   $f_1$(1285)     &         $K_3^*$(1780)               &  \pmb{$\Delta$(1620)} &  \pmb{$\Sigma$}\\
   $\eta$(1295)    &         $K_2$(1820)                 &  \pmb{$\Delta$(1700)} &  \pmb{$\Sigma$(1385)}\\
   $f_1$(1420)     &         $K_3$(2320)     &          $\Delta$(1950) &            $\Sigma$(1940)\\
   ...             &         ...             &                       & ... \\
   $f_2$(1430)     &         $D $            &                       &            \pmb{$\Xi$}\\
   ...             &         ...             &                       & ... \\
   $\eta$(1475)    &         $D_0^*$(2400)   &                       &            \pmb{$\Omega$}\\
   ...             &         ...             &                       & ... \\
   $f_2$(2340)     &         $\Upsilon$(11020)&            &          $\Xi_b$\\
%
    \hline
    \hline
  \end{tabular}
}
  \caption{List of hadrons incorporated in the THESEUS of Ref.~\cite{Batyuk:2016qmb}, 
	i.e.\ THESEUS-v1. The resonances used in 3FD simulations \cite{Ivanov:2005yw} are marked by bold font. 
    \label{tab:hadrons}}
\end{table}

The 3FD list is used for the hadron-gas EoS in terms of 
which all the freeze-out densities (baryon, strange and energy ones) are transformed into the 
corresponding baryon, strange chemical potentials and temperature. 
In the 3FD model, all the available energy and 
baryon number is distributed between the hadrons from the 3FD hadron list, except for the {$\Xi$} and {$\Omega$} hyperons. 
The latter two are just calculated with the deduced chemical potentials and temperature, and do not participate
in the balance of conserved quantities because of their negligible multiplicity as 
compared with that of other strange hadrons included in the 3FD table. 
Any additional (to the 3FD table) hadronic resonance, sampled in THESEUS, brings an excess
energy and baryon charge with respect to the 3FD baseline. 
This excess violates the energy, baryon number and strangeness conservation. 
If we use the original THESEUS-v1 table of hadronic resonances, it overestimates 
the total energy and baryon charge and produces non-zero net strangeness with respect to the pure 3FD calculation. 
Therefore, in the updated version, i.e. THESEUS-v2, we reduce the 
list of hadronic resonances precisely to the 3FD one in order for the average energy, strangeness and baryon charge of the produced events to be consistent with the total energy, strangeness and baryon charge of the underlying fluids in 3FD.

In principle, an alternative way to correct the conservation issue is possible. 
The local temperature, baryon and strange chemical potentials, 
provided by the 3FD input, could be recalculated
proceeding from the respective conservation laws in the extended table of hadronic resonances. This alternative way also has certain shortcomings. 
We would have to deal with a large number 
of resonances with not well-known decay modes. 
This would result in some uncertainties in the numbers and spectra of produced stable particles. 
Though, we expect that the final yields and spectra of produced stable particles would only insignificantly change as compared to the 3FD case because approximately the same number stable particles 
(because of baryon-number, strangeness and energy conservation) 
will originate from a larger variety of primordial resonances. 
Of course, the above-mentioned uncertainties
in decay modes of heavy resonances would also contribute to the change of these numbers and spectra.

In the example of the Au+Au collisions at~$\sqrt{s_{NN}}=$~9.2~GeV
analysed sect.~\ref{Rapidity distributions}, 
the original extended THESEUS table of hadronic resonances 
(THESEUS-v1) overshoots the original 3FD proton distribution by approximately~10\%.
This THESEUS-v1 excess of protons demonstrates the above mentioned effect of the excess baryon charge.  
The effects of the resonance table on~$K^+$ and~$K^-$ 
are different in relative values: approximately~10\% for~$K^-$ and~5\% for~$K^+$.
However, they are equal in absolute values 
because of strangeness conservation.  
The effect of the resonance table is stronger at high collision energies 
because a larger number of hadronic resonances beyond the 3FD list gets 
noticeably excited. At~$\sqrt{s_{NN}}<$ 5~GeV the effect of the resonance table
is negligible, it amounts to less than~1\% in terms of the midrapidity proton density.

\section{Recalculation of the baryon chemical potential} 
\label{Recalculation}

In order to implement the thermodynamic 
approach of light-nuclei formation, we first should recalculate the baryon chemical 
potential, proceeding from the local baryon number conservation
   \begin{eqnarray}
   \label{B-3FD}
 n_{{\rm primordial}\;N} (x;\mu_B,T) &+& \sum_{\rm hadrons} n_i(x;\mu_B,\mu_S,T) 
\cr
=
n_{{\rm observable}\;N}(x;\mu'_B,T) &+& \sum_{\rm hadrons} n_i(x;\mu'_B,\mu_S,T) 
\cr
&+& \sum_{\rm nuclei} n_c(x;\mu'_B,\mu_S,T).
   \end{eqnarray}
The sum over ``hadrons'' runs over the list of hadrons incorporated in the 3FD
(bold entries in Tab. \ref{tab:hadrons}) excluding the nucleon ($N$), the density of which 
is presented explicitly. 
The sum over ``nuclei'' runs over the list of light nuclei 
in Tab. \ref{tab:clusters}.
$n_i$ and $n_c$ are the local baryon densities of~$i$-th species of hadron and~$c$-th nucleus respectively, which 
depend on the local baryon~($\mu_B,\mu'_B$) and strange~($\mu_S$) chemical potentials 
and the temperature ($T$) at the freeze-out hypersurface. 
$\mu_B$ is the baryon chemical potential 
in terms of the primordial nucleons provided by the 3FD input,  
$\mu'_B$ is that in terms of the observable nucleons.

In fact, the recalculation of the baryon chemical potential also affects the energy and strangeness conservations. 
However, we do not additionally tune 
$\mu_S$ and $T$, because these conservations turn out to be quite good already: 
the total energy is conserved with the accuracy of 3\% and 
on average a few units of net strangeness are gained in the sampling
for the central Au+Au collision at $\sqrt{s_{NN}}=$~2.7~GeV. 
The total strangeness is not well conserved. However, it should be kept in mind 
that the strangeness production in the 3FD is poorly defined at this low energy. 
The strangeness produced at this collision energy should be reduced 
by a factor of~$\gamma_S\approx 0.2$~\cite{Ivanov:2013yqa}, 
which accounts for additional strangeness suppression due to constraints of the canonical ensemble \cite{Koch:1986ud}. 
This is the lowest explored collision energy, 
at which there is the largest fraction of nucleons bound in light nuclei. 
With the increase of collision energy the accuracy of the energy and strangeness conservation becomes better. 
For central Pb+Pb collisions at 
$E_{\rm lab}=$~20$A$~GeV, the energy is conserved within 1\%. 
Note that the coalescence implemented in the 3FD also 
reduces the total energy of the system because of a reduction of the number of degrees of freedom.

\section{Isotopic content of produced hadrons} 
\label{Isotopic} 

In the 3FD model, particles are not isotopically distinguished, i.e. the model deals with nucleons, pions, etc.
rather than with protons, neutrons, $\pi^+$, $\pi^0$, $\pi^-$, etc. 
Therefore, all quantities proportional to the number of protons are calculated for nucleons 
(after decays of all resonances) 
and then are scaled with the factor $(Z_p+Z_t)/(A_p+A_t)$, where $Z_\alpha$ is the charge of the colliding nucleus 
($\alpha=$ {\bf p}rojectile or {\bf t}arget) and~$A_\alpha$ is its mass number. 
This is just an approximate recipe to estimate the difference between proton and neutron yields. 
Other species (pions, kaons, etc.) are equally distributed between isotopic states, e.g., numbers of pions 
and kaons are~${N_{\pi^+}=N_{\pi^-}=N_{\pi^0}=N_{\pi}/3}$ and~${N_{K^+}=N_{K^0}=N_{K}/2}$, respectively.

THESEUS distinguishes different isospin states in the multiplets. However, by default it simulates isospin-symmetric matter because 
no information on its isotopic content is available from the 3FD. 
Nevertheless, even in isospin-symmetric matter the multiplicities of~$\pi^\pm$ and~$\pi^0$ 
and those of~$K^\pm$ and~$K^0$
differ because of the small differences in their masses~\cite{pdg2005} which are taken into account in THESEUS. 
If the weak decays are allowed, which is an option in THESEUS-v2, 
the difference between~$\pi^\pm$ and~$\pi^0$ becomes larger. 
The above points, of course, do not exhaust the full list of reasons resulting in the isotopic difference.  
This prescription is a good approximation at high collision energies, while at low energies it ignores the difference in yields of different isotopic species.

For the protons, THESEUS-v2 uses a slightly improved 3FD recipe. 
The fraction of protons is calculated as 
   \begin{eqnarray}
   \label{Rp}
 R_{\rm proton} =
\frac{Z_{\rm participants}- N_{d} - N_{t} - 2N_{^3\text{He}} - 2N_{^4\text{He}}}%
{B_{\rm participants} - 2N_{d} - 3N_{t} - 3N_{^3\text{He}} - 4N_{^4\text{He}}}
\cr
   \end{eqnarray}
where $B_{\rm participants}$  and 
$Z_{\rm participants}=B_{\rm participants}(Z_1+Z_2)/(A_1+A_2)$
are the total baryon number and electrical charge of participants, $N_{\rm nucleus}$ is the multiplicity of the produced light nucleus. $B_{\rm participants}$ is calculated within THESEUS-v2. 
Eq. (\ref{Rp}) takes into account that the fraction of observed protons is changed if some protons are bound in light nuclei.  
The figures \ref{fig:Au43b2}~(a), \ref{fig:Au2mix}
and \ref{fig:Pb20mix_i6} are produced precisely this way. 

The numbers of originally generated  tritons and $^3$He isotopes, $N_{t}$ and $N_{^3\text{He}}$, are equal because of the above mentioned  isotopic symmetry of the 3FD input. 
Therefore, these numbers, $N_{t}$ and $N_{^3\text{He}}$,  are also scaled with the factors 
$(N_p+N_t)/(A_p+A_t)$ and $(Z_p+Z_t)/(A_p+A_t)$, respectively, in order to take into 
account the initial isotopic imbalance of colliding nuclei.

\section{Coalescence in the 3FD} 
  \label{Coalescence}

The fragment production within the 3FD coalescence model~\cite{Ivanov:2005yw,Ivanov:2017nae} is described
similar to that it was done in Ref.~\cite{Russkikh:1993ct}. 
It is assume that $N$ neutrons and $Z$ protons, falling within a
6-dimensional phase volume
$(\frac{4}{3} \pi p_{NZ}^3) (\frac{4}{3} \pi r_{NZ}^3)$
at the freeze-out stage, form a $(N,Z)$-fragment. Here
$p_{NZ}$ and $r_{NZ}$ are the parameters of the coalescence
model, which are, in principle, different for different
$(N,Z)$-fragments.
The consideration below concerns a single cell in the configuration space. 
To avoid multiple subscripts in the notation we suppress the cell subscript. 
We calculate the distribution of
observable~$(N,Z)$-fragments as follows
%
\begin{eqnarray}
\label{dN(NZ)/dp(a)}
&&E_A \; \frac{d^3 \tilde{N}_{N,Z}}{d^3 P_A} =
\frac{N_{\text{tot}}^N Z_{\text{tot}}^Z}{A_{\text{tot}}^A} \;
A \;
\frac{(\frac{4}{3} \pi p_{NZ}^3 / M_N)^{A-1}}{N! Z!} \;
\cr
&&\times 
\left(
\frac{V_{NZ}}{V}
\right)^{A-1} \;
\left(
E \; \frac{d^3 \tilde{N}^{(N)}}{d^3 p}
\right)^{A}, 
\end{eqnarray}
%
where 
$d^3 \tilde{N}^{(N)}/d^3 p$ is the distribution of observable nucleons. Here
$N_{\text{tot}}=N_p+N_t$,
$Z_{\text{tot}}=Z_p+Z_t$ and
$A_{\text{tot}}=A_p+A_t$
are the total numbers of neutrons, protons and nucleons in the
projectile-plus-target nuclei, respectively, $A=N+Z$,
$E_A=AE$,
${\bf P}_A=A{\bf p}$, 
$V_{NZ}=\frac{4}{3} \pi r_{NZ}^3$,
and $M_N$ is the nucleon mass. 
$V=\bar{A}_{\text{cell}}/n_c$ is the total volume of the frozen-out cell,
where $n_c$ is the freeze-out baryon density and
%
\begin{equation}
\label{A(tot)}
\bar{A}_{\text{cell}} =
\int d^3 p
\; \frac{d^3 N^{(N)}}{d^3 p}
\end{equation}
%
is the total number of primordial participant nucleons.
Here we denote the distributions of observable (i.e. after the
coalescence) nucleons and fragments by a tilde
sign, in contrast to the primordial nucleon distribution.
Defining a new parameter
%
\begin{equation}
\label{P(NZ)}
P_{NZ}^3 =
\frac{4}{3} \pi p_{NZ}^3 \;
V_{NZ} \; n_c \;
\left(
\frac{A}{N! Z!} \;
\right)^{1/(A-1)},
\end{equation}
%
we can write down~eq.~(\ref{dN(NZ)/dp(a)}) in a simpler form
%
\begin{equation}
\label{dN(NZ)/dp(b)}
E_A \; \frac{d^3 \tilde{N}_{N,Z}}{d^3 P_A} =
\frac{N_{\text{tot}}^N Z_{\text{tot}}^Z}{A_{\text{tot}}^A} \;
\left(
\frac{P_{NZ}^3}{M_N \bar{A}_{\text{cell}}}
\right)^{A-1}
\left(
E \; \frac{d^3 \tilde{N}^{(N)}}{d^3 p}
\right)^{A},
\end{equation}
%
where 
$d^3 {N}^{(N)}/d^3 p$ is the distribution of
observable nucleons, i.e. those after the coalescence.  
In this form the fragment distribution contains only a single phenomenological 
parameter,  $P_{NZ}$, that defines the total normalization of the distribution. 
These equations for different~$N$ and~$Z$ form a set of equations,
since the nucleon distribution in the r.h.s. is an observable
distribution rather than a primordial one. To make this system
closed, one should add a condition of the baryon  number conservation
%
\begin{equation}
\label{N-conserv.}
E \; \frac{d^3         N^{(N)}}{d^3 p} =
E \; \frac{d^3 \tilde{N}^{(N)}}{d^3 p} +
\sum_{N,Z \; (A>1)} \; A^3 \;
E_A \; \frac{d^3 \tilde{N}_{N,Z}}{d^3 P_A}.
\end{equation}
%
Thus calculated distribution of observable fragments is summed over all cells 
in order to obtain the total momentum distribution of fragments.  
The~$P_{NZ}$ parameters are fitted to reproduce normalization of 
spectra of light fragments. 

%
\begin{table}[htb]
\begin{ruledtabular}
  \begin{tabular}{|cc|cc|}
$E_{\scr{lab}}$ &[$A\cdot$GeV]     &  2    & 20 \\ 
$P(\rm{d})$ &[MeV/c]               &  850  & 513\\ 
$P(\rm{t})$ &[MeV/c]               &  850  & 474\\ 
$P(^3\rm{He})$ &[MeV/c]            &  850  & 474\\ 
$P(^4\rm{He})$ &[MeV/c]            &  875  & 528\\ 
  \end{tabular}
\caption{Coalescence parameters, see~Eq.~(\ref{P(NZ)}),    
used in 3FD simulations of Au+Au (2$A$~GeV) and  Pb+Pb (20$A$~GeV)
collisions at various incident energies~$E_{\scr{lab}}$. 
}
\label{tab:2}
\end{ruledtabular}
\end{table}
Table \ref{tab:2} presents results of the fit of the~$P_{NZ}$ parameters 
made in~Ref.~\cite{Russkikh:1993ct} for Au+Au collisions and in Ref.~\cite{Ivanov:2017nae}
for the Pb+Pb data~\cite{Anticic:2016ckv}.


\newpage


\begin{thebibliography}{999}
  %
  \bibitem{Batyuk:2016qmb} P.~Batyuk {\it et al.},
  Phys.\ Rev.\ C {\bf 94}, 044917 (2016)
	[arXiv:1608.00965 [nucl-th]].
  %
  \bibitem{Ivanov:2005yw}
  Y.~B.~Ivanov, V.~N.~Russkikh and V.~D.~Toneev,
  Phys.\ Rev.\ C {\bf 73}, 044904 (2006)
	[nucl-th/0503088].
  %
\bibitem{Ivanov:2013wha}
Y.~B.~Ivanov,
Phys. Rev. C \textbf{87}, no.6, 064904 (2013)
[arXiv:1302.5766 [nucl-th]].
  %
  \bibitem{Bass:1993em} S.~A.~Bass, R.~Mattiello, H.~St\"ocker,
  W.~Greiner and C.~Hartnack,
  Phys.\ Lett.\ B {\bf 302}, 381 (1993).
  %
  \bibitem{Bass:1998ca} S.~A.~Bass {\it et al.},
  Prog.\ Part.\ Nucl.\ Phys.\ {\bf 41}, 255 (1998)
	[nucl-th/9803035].
	%
	\bibitem{Stephans:2006tg}
  G.~S.~F.~Stephans,
  J.\ Phys.\ G {\bf 32}, S447 (2006)
   [nucl-ex/0607030].
  %
  \bibitem{SPS-scan} 
  M.~Gazdzicki [NA49 and NA61/SHINE Collaborations],
  J.\ Phys.\ G {\bf 38}, 124024 (2011) 
[arXiv:1107.2345 [nucl-ex]].
%
  \bibitem{FAIR} 
T.~Ablyazimov \textit{et al.} [CBM],
Eur. Phys. J. A \textbf{53}, no.3, 60 (2017)
[arXiv:1607.01487 [nucl-ex]].
  %
  \bibitem{NICA}
  V.~D.~Kekelidze, V.~A.~Matveev, I.~N.~Meshkov, A.~S.~Sorin and G.~V.~Trubnikov,
  Phys.\ Part.\ Nucl.\  {\bf 48}, no. 5, 727 (2017).
  %
  \bibitem{gasEOS}
  V.~M.~Galitsky and I.~N.~Mishustin,
  Yad.\ Fiz.\ {\bf 29}, 363 (1979).
  %
  \bibitem{Toneev06}
  A.~S.~Khvorostukin, V.~V.~Skokov, V.~D.~Toneev and K.~Redlich,
  Eur.\ Phys.\ J.\ C {\bf 48}, 531 (2006)
   [nucl-th/0605069].
  %
\bibitem{Ivanov:2013yqa}
Y.~B.~Ivanov,
Phys. Rev. C \textbf{87}, no.6, 064905 (2013)
[arXiv:1304.1638 [nucl-th]].
  %
  %
  \bibitem{71} V. N. Russkikh, Yu. B. Ivanov,
  Phys. Rev. C {\bf 76}, 054907 (2007)
   [nucl-th/0611094].
  %
  \bibitem{74} Yu. B. Ivanov, V. N. Russkikh,
  Yad. Fiz.  {\bf 72}, 1288-1294 (2009)
	[arXiv:0810.2262 [nucl-th]].
%

\bibitem{pdg2005}
S.~Eidelman \textit{et al.} [Particle Data Group], Phys. Lett. B {\bf 592}, 1 (2004).
%
\bibitem{Abelev:2009bw}
B.~I.~Abelev \textit{et al.} [STAR],
Phys. Rev. C \textbf{81}, 024911 (2010)
[arXiv:0909.4131 [nucl-ex]].
%
\bibitem{Werner:2013tya}
K.~Werner, B.~Guiot, I.~Karpenko and T.~Pierog,
Phys. Rev. C \textbf{89}, no.6, 064903 (2014)
[arXiv:1312.1233 [nucl-th]].
%
\bibitem{Shuryak:2019ikv}
E.~Shuryak and J.~M.~Torres-Rincon,
Phys. Rev. C \textbf{101}, no.3, 034914 (2020)
[arXiv:1910.08119 [nucl-th]].
%
\bibitem{Ivanov:2017nae} 
  Y.~B.~Ivanov and A.~A.~Soldatov,
  Eur.\ Phys.\ J.\ A {\bf 53}, no. 11, 218 (2017)
[arXiv:1703.05040 [nucl-th]].
%
\bibitem{Koch:1986ud}
P.~Koch, B.~Muller and J.~Rafelski,
Phys. Rept. \textbf{142}, 167-262 (1986). 
%
\bibitem{Klay:2003zf}
J.~L.~Klay \textit{et al.} [E-0895],
Phys. Rev. C \textbf{68}, 054905 (2003)
[arXiv:nucl-ex/0306033 [nucl-ex]].
%
\bibitem{Russkikh:1993ct}
V.~N.~Russkikh, Y.~B.~Ivanov, Y.~E.~Pokrovsky and P.~A.~Henning,
Nucl. Phys. A \textbf{572}, 749-790 (1994). 
%
  \bibitem{Anticic:2016ckv} 
  T.~Anticic {\it et al.} [NA49 Collaboration],
  Phys.\ Rev.\ C {\bf 94}, no. 4, 044906 (2016)
[arXiv:1606.04234 [nucl-ex]].
%
\bibitem{Scheibl:1998tk}
R.~Scheibl and U.~W.~Heinz,
Phys. Rev. C \textbf{59}, 1585-1602 (1999)
[arXiv:nucl-th/9809092 [nucl-th]].
%
\bibitem{Liu:2019nii}
H.~Liu, D.~Zhang, S.~He, K.~j.~Sun, N.~Yu and X.~Luo,
Phys. Lett. B \textbf{805}, 135452 (2020)
[arXiv:1909.09304 [nucl-th]].
%
\bibitem{Zhu:2015voa}
L.~Zhu, C.~M.~Ko and X.~Yin,
Phys. Rev. C \textbf{92}, no.6, 064911 (2015)
[arXiv:1510.03568 [nucl-th]].
%
\bibitem{Dong:2018cye}
Z.~J.~Dong, G.~Chen, Q.~Y.~Wang, Z.~L.~She, Y.~L.~Yan, F.~X.~Liu, D.~M.~Zhou and B.~H.~Sa,
Eur. Phys. J. A \textbf{54}, no.9, 144 (2018)
[arXiv:1803.01547 [nucl-th]].
%
\bibitem{Sombun:2018yqh}
S.~Sombun, K.~Tomuang, A.~Limphirat, P.~Hillmann, C.~Herold, J.~Steinheimer, Y.~Yan and M.~Bleicher,
Phys. Rev. C \textbf{99}, no.1, 014901 (2019)
[arXiv:1805.11509 [nucl-th]].
%
\bibitem{Zhao:2020irc}
W.~Zhao, C.~Shen, C.~M.~Ko, Q.~Liu and H.~Song,
Phys. Rev. C \textbf{102}, no.4, 044912 (2020)
[arXiv:2009.06959 [nucl-th]].
%
\bibitem{Oliinychenko:2020ply}
D.~Oliinychenko,
Nucl. Phys. A \textbf{1005}, 121754 (2021)
[arXiv:2003.05476 [hep-ph]].
%
\bibitem{Donigus:2020fon}
B.~D\"onigus,
Eur. Phys. J. A \textbf{56}, no.11, 280 (2020).
%
\bibitem{Weil:2016zrk}
J.~Weil, V.~Steinberg, J.~Staudenmaier, L.~G.~Pang, D.~Oliinychenko, J.~Mohs, M.~Kretz, T.~Kehrenberg, A.~Goldschmidt and B.~B\"auchle, \textit{et al.}
Phys. Rev. C \textbf{94}, no.5, 054905 (2016)
[arXiv:1606.06642 [nucl-th]].
%
\bibitem{Oliinychenko:2018ugs}
D.~Oliinychenko, L.~G.~Pang, H.~Elfner and V.~Koch,
Phys. Rev. C \textbf{99}, no.4, 044907 (2019)
[arXiv:1809.03071 [hep-ph]].
%
\bibitem{Aichelin:2019tnk}
J.~Aichelin, E.~Bratkovskaya, A.~Le F\`evre, V.~Kireyeu, V.~Kolesnikov, Y.~Leifels, V.~Voronyuk and G.~Coci,
Phys. Rev. C \textbf{101}, no.4, 044905 (2020)
[arXiv:1907.03860 [nucl-th]].
%
\end{thebibliography}
\end{document}